# A massive binary black-hole system in OJ287 and a test of general relativity

M. J. Valtonen[1], H. J. Lehto[1], K. Nilsson[1], J. Heidt[2], L. O. Takalo[1], A. Sillanpää[1], C. Villforth[1], M. Kidger[3], G. Poyner[4], T. Pursimo[5], S. Zola[6,7], J.-H. Wu[8], X. Zhou[8], K. Sadakane[9], M. Drozdz[7], D. Koziel[6], D. Marchev[10], W. Ogloza[7], C. Porowski[6], M. Siwak[6], G. Stachowski[7], M. Winiarski[6], V.-P. Hentunen[11], M. Nissinen[11], A. Liakos[12], S. Dogru[13]

*1 Department of Physics and Tuorla Observatory, University of Turku, Vaisäläntie 20, FI-21500 Piikkiö, Finland*

*2 Landessternwarte Heidelberg, Königstuhl, 69117 Heidelberg, Germany*

*3 Herschel Science Centre, European Space Astronomy Centre, European Space Agency, Villafrance del Castillo Satellite Tracking Station, P.O.Box 78, 28691 Villanueva de la Canada, Madrid, Spain, and INSA, Paseo del Pintor Rosales 34, 28008 Madrid, Spain*

*4 British Astronomical Association Variable Star Section, 67 Ellerton Road, Kingstanding, Birmingham B44 0QE, England*

*5 Nordic Optical Telescope, Apartado 474, E-38700 S/C de La Palma, Spain*

*6 Astronomical Observatory of the Jagiellonian University, u1. Orla 171, 30-224 Cracow, Poland*

*7 Mt.Suhora Observatory, Pedagogical University, ul. Podchorazych 2, 30-084, Cracow, Poland*

*8 National Astronomical Observatories, Chinese Academy of Sciences, 20A Datun Road, Beijing 100012, China*

*9 Astronomical Institute, Osaka-Kyoiku University, Asahigaoka, Kashiwara, Osaka 582-8582, Japan*

*10 Department of Physics, Shoumen University, 9700 Shoumen, Bulgaria*

*11 Warkauden Kassiopeia ry, Härkämäentie 88, 79480 Kangaslampi, Finland*

*12 Department of Astrophysics, Astronomy and Mechanics, Faculty of Physics, University of Athens, Panepistimiopolis, GR-15784 Zografos, Athens, Greece*

*13 Canakkale Onsekiz Mart University, Faculty of Physics, TR-17020 Canakkale, Turkey*

**Tests of Einstein's general theory of relativity have mostly been carried out in weak gravitational fields where the space-time curvature effects are first-order deviations from Newton's theory[1–6]. Binary pulsars[4] provide a means of probing the strong gravitational field around a neutron star, but strong-field effects may be best tested in systems containing black holes[7,8]. Here we report such a test in a close binary system of two candidate black holes in the quasar OJ287. This quasar shows quasi-periodic optical outbursts at 12 yr intervals, with two outburst peaks per interval[9,10]. The latest outburst occurred in September 2007, within a day of the time predicted by the binary black-hole model and general relativity[11]. The**

**observations confirm the binary nature of the system and also provide evidence for the loss of orbital energy in agreement (within 10 per cent) with the emission of gravitational waves from the system[12]. In the absence of gravitational wave emission the outburst would have happened twenty days later[13].**

The quasar OJ287 has been observed optically since the late 19th century, but was identified as a quasar only in 1968. This quasar is of great interest because it shows a quasi-periodic pattern of prominent outbursts; these consist of 11 well-recognized outbursts and several probable outbursts, all observed before 2007[14] (see the [Supplementary Information](#) for the full light curve). Because the outbursts seem to come in pairs separated by one to two years, and the pairs occur about 12 years apart, we proposed a model in which a secondary body (a black hole) pierces the accretion disk of the primary black hole and produces two impact flashes per period[10]. Alternative models attribute the observed behaviour to oscillations in an accretion disk[15] or a jet[16] of a single black hole, or to variations of the accretion rate in a disk[9] or a wobble of a jet[17] in a binary black hole system. However, these processes do not produce the sharp flashes that we consider here[14]. The consequences our model has for jet emission, the primary emission mode, have been calculated in several papers[14,18]. Here we concentrate on the optical flashes associated with disk impacts. We treat them as time signals that tell us how the secondary moves around the primary in its near-keplerian orbit.

Apart from the 12-year quasi-periodicity, the most prominent bright outbursts do not repeat themselves in any easily recognizable pattern. However, the following mathematical formula gives the times of all the outburst peaks observed during the twentieth century[10]: take a keplerian orbit with post-newtonian corrections and assume that an outburst happens at a constant phase angle as well as at the same phase angle plus 180 degrees. This provides a well-defined mathematical series of outburst times. Although the series cannot be represented by a closed-form mathematical function (the same is true for the original Kepler problem), the times in the series are quickly calculable by computer. If the orbit is eccentric, there is a pair of outburst times that are close to each other relative to the orbital period.

Assuming that the constant phase angle is provided by the secondary passing through the accretion disk of the primary, the orbit can be reconstructed by using all the available information on past outbursts and straightforward astrophysics. For any general sequence of quasi-random times, there is unlikely to be a solution. There is an algorithm that converges to a solution of orbital elements if a solution exists[13]. Non-precessing orbits can be rejected immediately; no solution exists for such orbits. This is promising because, even at first order in general relativity, the major axis of the orbit must precess. Deviations of as little as a few weeks from the observed outburst times also result in insolubility. Thus, the observed sequence of outburst times is quite particular, and unlikely to arise by chance, as has been demonstrated quantitatively[19].

The complete solution of the problem requires a light curve coverage of six well-defined outbursts. When the method was first applied[10] there existed only four such peaks. It was not until the early 2007 that there were enough data to calculate a definite orbit[11]. The precession rate of the major axis of this orbit is 39.0 degrees per orbit, the eccentricity of the orbit is 0.663, and the mass of the primary black hole is $18.0 \times 10^9$ solar masses. These values are reasonable: merging binaries are expected to have eccentricities similar to this at intermediate stages of evolution[20], and the mass of the black hole is at the upper end of the mass range in quasars[21] (which is encouraging, as OJ287 is among the brightest quasars).

The first test of the orbit solution took place in September 2007, when the second one of the double peaks was due to appear. The date predicted by the model was September 13 plus or minus two days. One of the sources of uncertainty in the prediction comes from the zero-point error in timing the beginning of the 1983 outburst, generally taken to be 1983.00. This figure is the fiducial time which is used in the models to compare the beginning times of different outbursts. On the basis of a light curve fit such as that shown in Fig. 1, this error is about one day. A similar error arises from the variation of orbital parameters within the range allowed by the relative timing of different outbursts. A third source of error is related to uncertainties in the model, in particular to the effects of accretion disk bending[13]. This effect is small for the 2007 impact, being also of the order of one day. Thus, the various sources of error contribute an uncertainty of about two days in the timing of the 2007 optical flash.

The date of September 13 is important in other respects also: if the system does not lose energy by emission of gravitational radiation, and thus the orbit does not shrink, the outburst times are delayed. In 2007 September the delay between an orbit of constant semi-major axis and an orbit shrinking by emission of gravitational radiation was 20 days; that is, the two solutions of the post-newtonian Kepler equation differed by this amount. This is a test of general relativity that was first suggested in 1997[12] ; and can now be performed.

The difference between the two cases would be easily resolvable in monitoring observations if there were enough observers in different parts of the world to cover the nights missed owing to unpredictable circumstances at any one observatory. The main difficulty was the actual season; the outburst could have started any time after September 11. Because OJ287 undergoes conjunction with the Sun only a month before this, it is a very difficult object to observe during this season. Moreover, many telescopes do not turn close enough to the horizon to start monitoring OJ287 in early September. Here we report results of OJ287 monitoring between 2007 September 4 and 2007 October 20, 2007.

A total of 100 photometric observations of OJ287 were obtained between 2007 September 4 and 2007 October 20, as shown in Fig.2. We see that there were several outbursts from OJ287 during this period. This is as expected, on the basis of calculations of the light curve for the 2005-2007 outburst season[18]. The 2005 outburst was both observed and predicted to be somewhat isolated[22,23], whereas the 2007 outburst was expected to be embedded in a series of outbursts of about equal brightness. The impact outburst we discuss here was expected to be the first outburst in the sequence, but not necessarily the brightest. For the purposes of our model it is important to identify the correct outburst in September 2007, by comparing the outbursts both with each other and with previous outbursts of the same nature.

The outburst beginning at September 12 was the brightest recorded during our period of observation. Thus, of those outbursts early in the sequence, this is the most promising one. One outburst was observed earlier, reaching its maximum brightness around September 8, and is clearly smaller than the outburst of September 12. However, maximal brightness alone is not a sufficient criterion to identify the impact outburst.

Another discriminant is the detailed light curve; the timescale of the outburst was calculated to be three days at the half-maximum, and there was a -3/2 power-law decay in brightness due to adiabatic energy losses[10]. At the impact of the secondary on the accretion disk, a hot bubble of gas is extracted from the disk[24] and expands at its internal sound speed. Once the bubble has expanded to the point at which it becomes optically thin, there is a burst of radiation. The radiation fades away with further expansion of the bubble. The time of rise to the maximum flux can be very rapid near the peak, as it is related to the travel time of light through the radiating region when it becomes fully transparent. The September 12 outburst fits this description very well, whereas the other outbursts do not. Figure 1 illustrates the light curves of major 1983, 2005 and 2007 outbursts, assuming that the September 12 outburst is identified as the impact flash.

A third identifying characteristic of the impact outburst involves polarization data. The 1983 outburst is the only one for which polarization measurements have been made frequently throughout the whole outburst event, from the flux rise to its decay. In 1983 the degree of polarization dropped rapidly during the flux rise[25]. This is also expected theoretically[14]. In contrast, the polarization usually rises during the outburst, as our light curve for October 10 – 16 demonstrates. As Fig. 2 shows, the September 12 outburst was most similar to the 1983 outburst in this respect also.

As evidence for gravitational waves in the OJ287 binary we note that the model without gravitational waves predicts an outburst in 2007 that takes place about 20 days later than it does in the standard model. As the outburst occurred as expected, we must conclude that gravitational waves are the probable cause of the shortening of the orbital period, and that the strength of the radiation is at the expected level with a timing accuracy of about 2 days out of 20, that is, the energy flux does not deviate from the value calculated using general relativity by more than 10%. In principle one could improve the prediction once future outbursts have been observed and the orbit model further improved.

However, if gravitational radiation were to follow general relativity but the second-order precession term were missing, the outburst would occur ten days earlier than was observed. Thus, the space-time curvature appears at the second order as general relativity predicts. The precision of this test is two days out of 10, that is, 20%. The next major periodic outburst is expected in early January 2016, by which time there may be methods to measure the gravitational waves directly.

26. Fiorucci, M. & Tosti, G. VRI photometry of stars in the fields of 12 BL Lacertae objects. Astron.&Astrophys.Suppl. 116, 403-407 (1996).

**Supplementary Information** is linked to the online version of the paper at www.nature.com/nature.

**Acknowledgements** This work is supported by the European community (project ENIGMA), the German Science Foundation, Japanese Ministry of Education, Culture, Sports, Science and Technology, Chinese Academy of Sciences and the Chinese National Natural Science Foundation, Finnish Society of Sciences and Letters, Finnish Academy of Science and Letters, Jenny and Antti Wihuri Foundation, Development Association Mansikka ry, and the municipality of Varkaus.

**Author contributions** M.J.V. and H.J.L. were responsible for the interpretation of the data, K.N. and J.H. organized the observational fieldwork, and other authors contributed data points.

**Author information** Reprints and permissions information is available at www.nature.com/reprints. Correspondence and requests for materials should be addressed to M.J.V. (mvaltonen2001@yahoo.com).

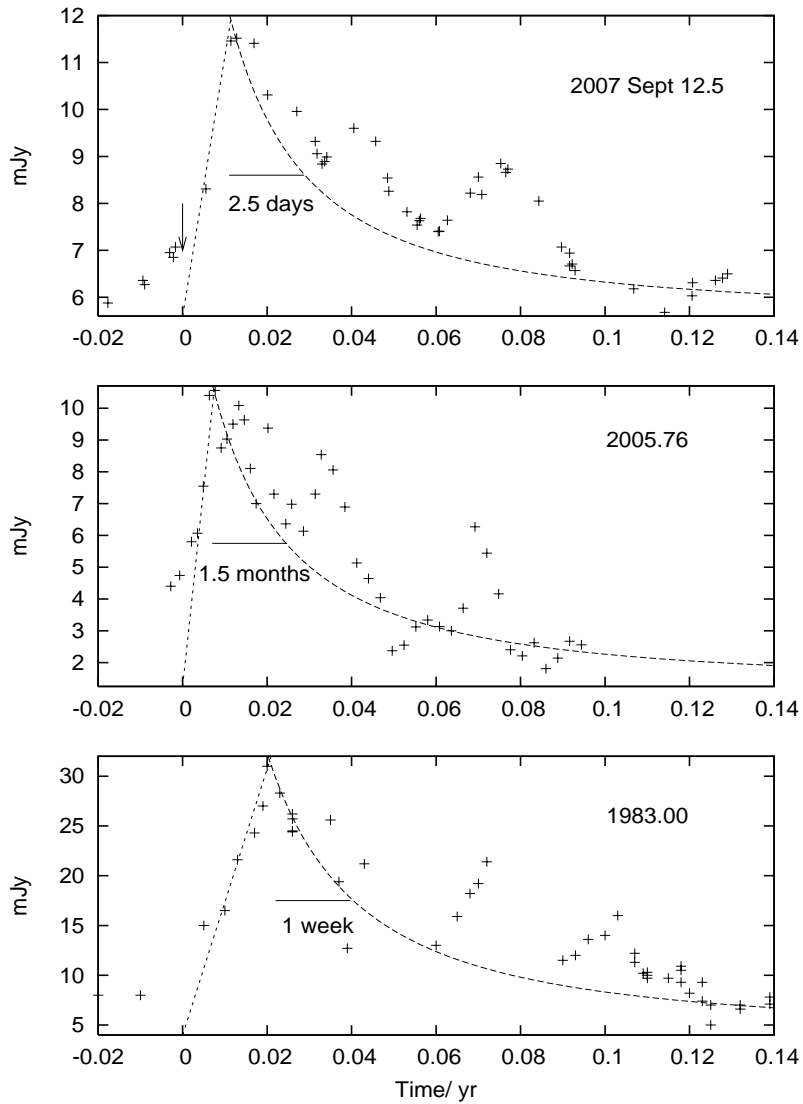

**Figure 1 Optical V-band fluxes of OJ287 vs. time in relative units.** The dashed lines show the theoretical predictions[10]. The horizontal axis displays time in years for 1983; the zero point corresponds to 1983.00. The time given in the upper right-hand corner of each panel indicates the nominal starting time of the outburst, that is, when the dashed line crosses the time axis. For the 2007 outburst the time scale has been stretched by a factor of 2.65, and for the 2005 outburst, contracted by a factor of 7.16, so that the intrinsically different decay timescales can be shown on the same scale. The timescales are different because the speed of impact, and consequently the internal sound speed of the radiating bubble, varies with the impact distance. The impact distance from the primary black hole was 13,300 AU in 2005, 4,800 AU in 1983 and 3,400 AU in 2007. The flux scales are given in mJy on the vertical axis. The arrow in the top panel points to the inferred origin of the outburst at 12:00 UT on 2007 September 12. The next observation after this was taken at 6:00 UT on September 13, close to the predicted starting time of the outburst. In each panel, the horizontal line segment indicates the decay timescale.

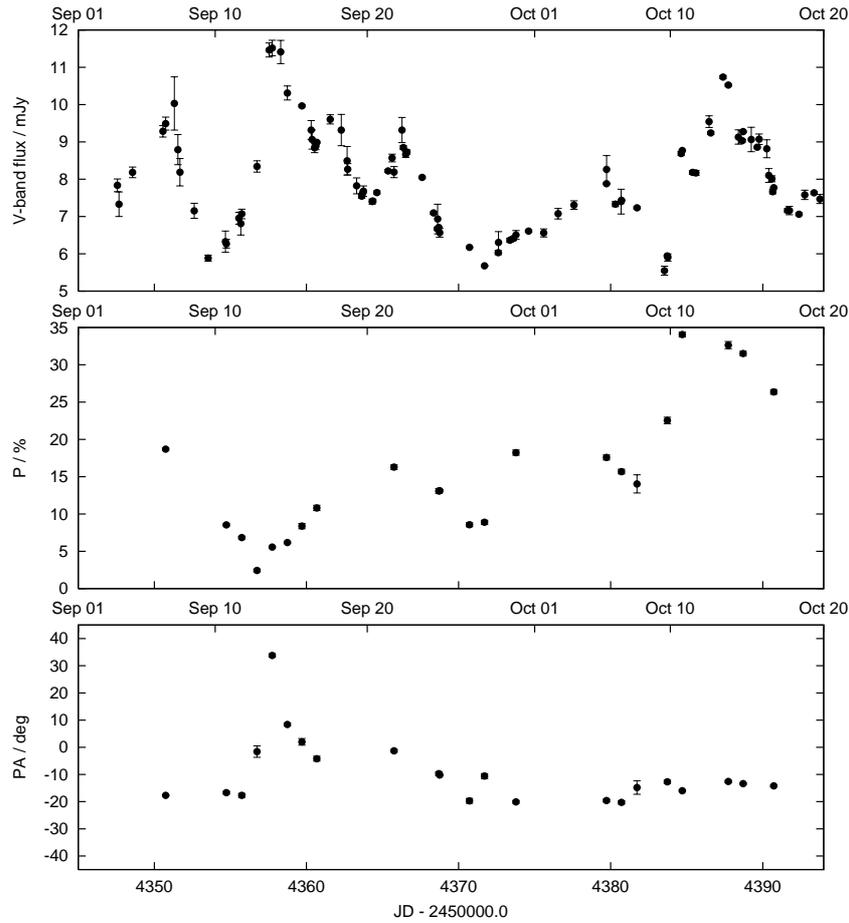

**Figure 2 Optical photometry and polarimetry of OJ287 in September-October, 2007.** The top panel shows data transformed to the V band, whereas the polarization data in the two lower panels are in the R band. The majority (93) of the 100 observations were made in the R band using 11 optical telescopes, of aperture sizes of 30–256 cm, in the Canary Islands, Europe, China and Japan. The observations were made in differential mode, that is, by obtaining charged-coupled-device images of OJ 287 and comparing the object's brightness to calibrated comparison stars in its field. The comparison sequence in ref. 26 was used, with star four as the primary reference. Star ten was used to check for offsets between different telescopes, caused, for example, by a non-standard filter band pass. In some cases offsets were found, but they were very small (<0.03 mag) and the magnitudes of OJ 287 have been corrected for this offset. The magnitudes were transformed to the V band using V-R = 0.4. Finally, the V-band magnitudes we converted to linear fluxes using the formula $F = 3{,}640 \cdot 10^{-0.4 \cdot V}$ Jy. We also collected 21 polarimetric observations in the R band using the 2.56-m Nordic Optical Telescope in La Palma and the Calar Alto 2.2-m telescope in Spain. All data were obtained using instruments employing a half-wave plate to rotate the direction of polarization and a calcite crystal to separate the incoming light into ordinary and extraordinary beams. This effectively cancelled out any instrumental effects and the background polarization. The instrumental polarization was checked using zero-polarization stars and was found to be negligible. The polarization angle was calibrated using highly polarized stars, with published polarization values. The s.e.m. error is shown by the error bars. Often it is smaller than the representative point.

**Supplementary information: The historical light curve of OJ287 in V band**

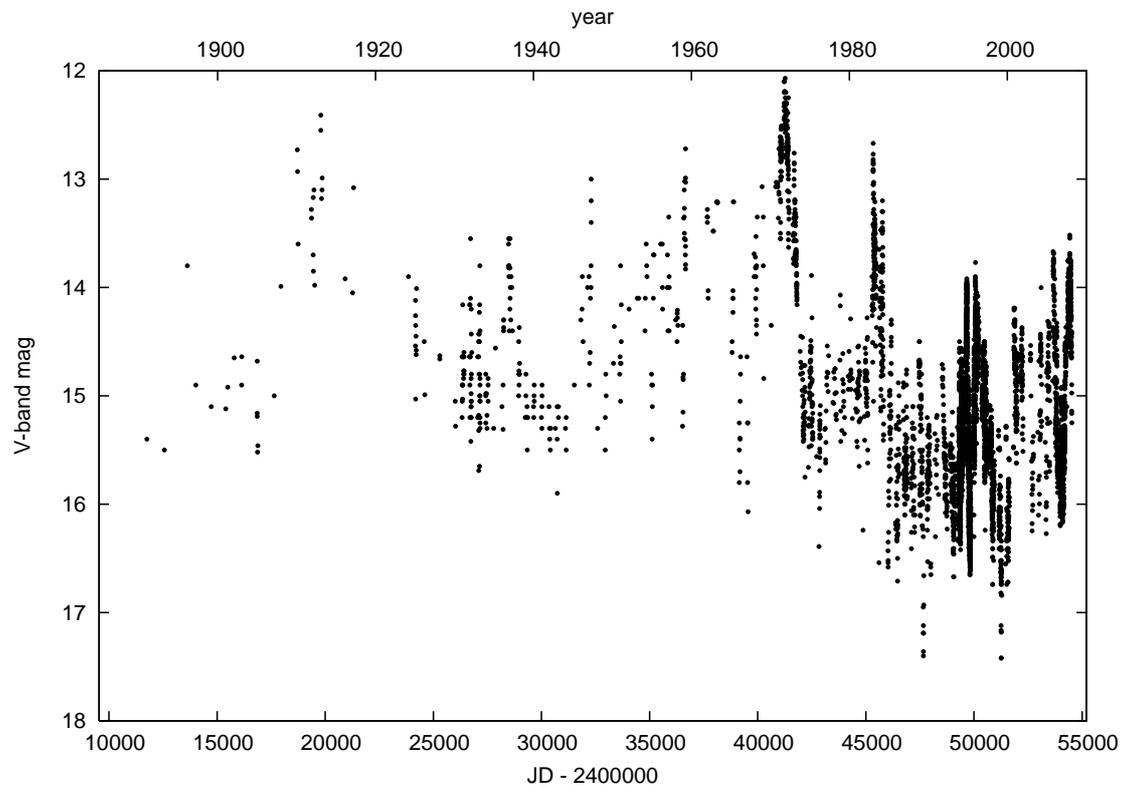